\documentclass[aps,prc,twocolumn,superscriptaddress,showpacs,floatfix,nofootinbib]{revtex4-1}
\usepackage{amsmath,graphicx}
\begin{document}

\title{Isolating dynamical net-charge fluctuations}

\author{Rudolph Rogly}
\affiliation{
Institut de physique th\'eorique, Universit\'e Paris Saclay, CNRS,
CEA, 91191 Gif-sur-Yvette, France}
\affiliation{MINES ParisTech, PSL Research University, 60 Boulevard Saint-Michel, 75006 Paris, France}
\author{Giuliano Giacalone}
\affiliation{
Institut de physique th\'eorique, Universit\'e Paris Saclay, CNRS,
CEA, 91191 Gif-sur-Yvette, France}  
\author{Jean-Yves Ollitrault}
\affiliation{
Institut de physique th\'eorique, Universit\'e Paris Saclay, CNRS,
CEA, 91191 Gif-sur-Yvette, France} 
\date{\today}

\begin{abstract}
  We modify the usual definitions of cumulants of net-charge fluctuations in a way that isolates dynamical fluctuations.
  The new observables, which we call dynamical cumulants, are robust with respect to trivial correlations induced by volume fluctuations and global charge conservation.
  We illustrate the potential of dynamical cumulants by carrying out Monte Carlo simulations where all correlations are trivial. 
The results of our simulations agree well with Relativistic Heavy Ion Collider (RHIC) data, and are used to illustrate that dynamical cumulants consistently isolate dynamical fluctuations.
\end{abstract}

\maketitle
\section{Introduction}

%``Statistical and dynamical parts of the cumulants of conserved charges in relativistic heavy ion collisions,''

One of the motivations for studying nucleus-nucleus collisions is to obtain information about the phase diagram of dense quantum chromodynamics (QCD)~\cite{Fukushima:2010bq}.
The conjectured existence of a critical point, although not yet confirmed by ab-initio calculations~\cite{Bazavov:2017dus}, has triggered experimental studies of particle number fluctuations~\cite{Stephanov:1998dy,Stephanov:2008qz}. 
Fluctuations of conserved charges (baryon number, electric charge, strangeness) are in fact interesting even if there is no critical point, because they can be computed from first principles in lattice QCD~\cite{Cheng:2008zh,Bazavov:2012vg}. 
This has motivated detailed analyses of fluctuations of the net proton number~\cite{Adamczyk:2013dal}, the net electric charge~\cite{Abelev:2012pv,Adamczyk:2014fia,Adare:2015aqk} and the net strangeness~\cite{Adamczyk:2017wsl}, hereafter generically referred to as net-charge fluctuations, which have been compared with lattice QCD results~\cite{Gupta:2011wh,Borsanyi:2013hza,Borsanyi:2014ewa,Borsanyi:2018grb} (see \cite{Luo:2017faz} for a recent review). 

The standard observables for net-charge fluctuations are cumulants, which are measured in experiment up to order 4~\cite{Aggarwal:2010wy}, and calculated on the lattice~\cite{Cheng:2008zh}.
Experimental results for cumulants of order 3 and 4 deviate from trivial expectations based on Poisson fluctuations~\cite{Adare:2015aqk}.
However, it has been realized that such deviations are also produced by uninteresting fluctuations referred to as non-dynamical, which have two distinct origins: 
Impact parameter fluctuations (also called volume fluctuations)~\cite{Skokov:2012ds,Braun-Munzinger:2016yjz,Bzdak:2018uhv,Pan:2018ryx}, and global charge conservation~\cite{Bzdak:2012ab}. 

In this paper, we construct new quantities, dubbed {\it dynamical cumulants\/}, which can be measured in any experiment, and differ from the ordinary cumulants only by trivial self-correlation terms.
Dynamical cumulants generalize factorial cumulants~\cite{Mueller:1971ez,DeWolf:1995nyp,Bzdak:2012ab,Broniowski:2017tjq} by taking into account the correlation induced by global charge conservation. 
In Sec.~\ref{s:1particle}, we construct dynamical cumulants in the simple case where there is only one species of particles~\cite{Bzdak:2016jxo}. 
In Sec.~\ref{s:expressions}, we generalize to the realistic case when there are both negatively and positively charged particles, and we provide explicit expressions for the dynamical cumulants of net charge fluctuations. 
In Sec.~\ref{s:volume}, we show that dynamical cumulants are remarkably insensitive to impact parameter fluctuations.
In Sec.~\ref{s:resonance}, we discuss their sensitivity to resonance decays in the hadronic phase. 
In Sec.~\ref{s:mc}, we illustrate the advantage of dynamical cumulants over ordinary cumulants by carrying out  Monte Carlo simulations where the correlations are solely due to impact parameter fluctuations and global charge conservation. 
We show that this simulation reproduces the seemingly non-trivial values of cumulants measured at RHIC, while dynamical cumulants are smaller by orders of magnitude. 

\section{Construction of dynamical cumulants}
\label{s:1particle}

We start by discussing the simple case where there is only one species of particles~\cite{Bzdak:2016jxo}. 
We denote by $N$  the number of particles detected in an event. 
$N$ fluctuates event to event. 
The goal is to construct a set of quantities which isolate the dynamical information contained in these fluctuations. 
We first discuss the simple case of the variance, then generalize to higher-order cumulants. 

\subsection{Variance}

Fluctuations can be characterized by the series of cumulants of the distribution of $N$, whose general definition will be recalled below. 
The first two cumulants are simply the mean and the variance:
\begin{eqnarray}
\label{meanvariance}
\kappa_1&\equiv&\langle N\rangle,\cr
\kappa_2&\equiv&\langle N^2\rangle-\langle N\rangle^2,
\end{eqnarray}
where angular brackets denote an average over a large sample of collision events, typically in a narrow centrality class. 

These definitions are general and apply both to integer and real variables. 
For an integer variable, alternatively, one may replace $N^2$ by the number of pairs $N(N-1)$. 
The corresponding quantity is the {\it factorial\/} cumulant~\cite{DeWolf:1995nyp} of order 2, which we denote by $f_2$: 
\begin{equation}
\label{robustvariance}
f_2\equiv\langle N(N-1)\rangle-\langle N\rangle^2. 
\end{equation}
The first term is the average number of pairs, and the second term is the number of uncorrelated pairs. 
Hence the factorial cumulant isolates the correlated part of the variance, $\kappa_2$, by subtracting the trivial part, $\langle N\rangle$, corresponding to Poisson fluctuations~\cite{Pruneau:2002yf,Pan:2013xya}. 

Global conservation laws (conservation of momentum, energy, charge) give rise to correlations among particles. 
If particles were emitted independently, the sum of charges would not be exactly the same in every event. 
Correlations induced by global conservation laws are to some extent trivial and must also be subtracted in order to isolate the dynamical correlations~\cite{Borghini:2000cm,Bzdak:2012an}. 
In this simple case where there is only one species of particles, we mimic the effect of the conservation law by assuming that the total number of particles for the whole event is fixed to some value $N^{\rm tot}$. 
If there is no other correlation, then the distribution of the number of particles seen in the detector, $N$, follows a binomial distribution:
\begin{equation}
\label{binomial}
P(N)=\begin{pmatrix}N^{\rm tot}\cr N\end{pmatrix} \alpha^{N}(1-\alpha)^{N^{\rm tot}-N},
\end{equation}
where $\alpha$ denotes the probability for an emitted particle to be seen by the detector~\cite{Braun-Munzinger:2016yjz}, which depends on its acceptance~\cite{Rustamov:2017lio}  and efficiency. 
Throughout this paper, we assume that $\alpha$ is known for the detector used. We neglect its variation within a centrality bin, as well as non-binomial efficiency corrections~\cite{Bzdak:2016qdc,Nonaka:2018mgw}. 
The upper bound on $N$,  $N\le N^{\rm tot}$, reduces the variance of the distribution, which is $(1-\alpha)\langle N\rangle$ for the binomial distribution, instead of just $\langle N\rangle$ for the Poisson distribution. 
We define the dynamical cumulant $d_2$ by subtracting this contribution:
\begin{equation}
\label{dynamicalvariance}
d_2\equiv\langle N(N-1+\alpha)\rangle-\langle N\rangle^2. 
\end{equation}
The dynamical cumulant thus defined coincides with the factorial cumulant $f_2$, Eq.~(\ref{robustvariance}), in the limit $\alpha\to 0$, and with the ordinary cumulant $\kappa_2$, Eq.~(\ref{meanvariance}), in the limit $\alpha\to 1$. 
In practice, detectors at RHIC and LHC cover a small fraction of the total phase space, and $\alpha$ is significantly smaller than unity. The dynamical cumulant is typically much closer to the factorial cumulant than to the ordinary cumulant.  

\subsection{Higher-order cumulants}

We now explain how this construction can be generalized to cumulants of arbitrary order. 
This can be done systematically using the formalism of generating functions. 
The ordinary cumulant $\kappa_n$ is defined by the power series expansion of a generating function:
\begin{equation}
 \label{defkn}
 \ln \left\langle \exp\left(zN\right)\right\rangle
\equiv\sum_{n=1}^{\infty}\kappa_n\frac{z^n}{n!}.
\end{equation}
Expanding both sides of the equation to order $z^2$, one recovers Eq.~(\ref{meanvariance}). 
If one pushes the expansion up to order $z^4$, one obtains the next cumulants, which are the skewness, $\kappa_3$, and the kurtosis, $\kappa_4$: 
\begin{eqnarray}
  \label{kappa34}
\kappa_3&=&\langle (N-\langle N\rangle)^3\rangle,\cr
\kappa_4&=&\langle (N-\langle N\rangle)^4\rangle-3\kappa_2^2.
\end{eqnarray}
The factorial cumulants, $f_n$, are defined through a different generating function: 
\begin{equation}
 \label{deffn}
 \ln \left\langle (1+z)^N\right\rangle
\equiv\sum_{n=1}^{\infty}f_n\frac{z^n}{n!}.
\end{equation}
If one expands $(1+z)^N$ to order $z^n$, the coefficient is the number of $n$-plets. 
The factorial cumulant thus defined isolates the true $n$-particle correlation~\cite{Bzdak:2012ab} by enumerating the number of $n$-plets at every order.  
By expanding the left-hand side of Eq.~(\ref{deffn}) to order $z^2$, one recovers Eq.~(\ref{robustvariance}). 

Finally, we define the dynamical cumulant $d_n$ through the following generating function:
\begin{equation}
 \label{defdn}
 \ln \left\langle\left(1+\frac{e^{\alpha z}-1}{\alpha}\right)^N\right\rangle
\equiv\sum_{n=1}^{\infty}d_n\frac{z^n}{n!}.
\end{equation}
Expanding the left-hand side to order $z^2$, one recovers Eq.~(\ref{dynamicalvariance}).
For $\alpha=1$, $d_n$ coincides with $\kappa_n$ defined by Eq.~(\ref{defkn}). 
In the limit $\alpha\to 0$, $d_n$ coincides with $f_n$ defined by Eq.~(\ref{deffn}). 
Thus, dynamical cumulants interpolate between ordinary cumulants and factorial cumulants. 
We now show that dynamical cumulants of order $n\ge 2$ vanish for a binomial distribution. 
If the distribution of $N$ is given by Eq.~(\ref{binomial}), one obtains:
\begin{eqnarray} 
\label{dynbinomial}
\left\langle \left(1+\frac{e^{\alpha z}-1}{\alpha}\right)^N\right\rangle
&=&
\sum_{N=0}^{N^{\rm tot}}
P(N) \left(1+\frac{e^{\alpha z}-1}{\alpha}\right)^N\cr
&=&
\exp\left( \alpha N^{\rm tot}z\right).
\end{eqnarray}
Inserting this equation in Eq.~(\ref{defdn}), one obtains $d_1=\alpha N^{\rm tot}=\langle N\rangle$, and $d_2=d_3=\dots=0$. 
This justifies the definition (\ref{defdn}). 

Dynamical cumulants can be expressed as a function of ordinary cumulants using Eqs.~(\ref{defkn}) and (\ref{defdn}), through an appropriate change of variables: 
\begin{equation}
\label{dnversuskn}
\sum_{n=1}^{\infty}\frac{d_n}{n!}z^n=
\sum_{n=1}^{\infty}\frac{\kappa_n}{n!}\ln^n\left( 1+\frac{e^{\alpha z}-1}{\alpha}  \right).
\end{equation}
Expanding the right hand side up to order $z^4$, one obtains the following explicit expressions: 
\begin{eqnarray}
\label{d1234}
d_1&=&\kappa_1,\cr
d_2&=&\kappa_2-(1-\alpha)\kappa_1,\cr
d_3&=&\kappa_3-3(1-\alpha)\kappa_2+(1-\alpha)(2-\alpha)\kappa_1,\cr
d_4&=&\kappa_4-6(1-\alpha)\kappa_3+(1-\alpha)(11-7\alpha)\kappa_2\cr
&&-(1-\alpha)(6-6\alpha+\alpha^2)\kappa_1.
\end{eqnarray}
Thus the dynamical cumulant of order $n$ differs from the ordinary cumulant $\kappa_n$ only by terms which involve lower-order cumulants. 
These subtracted terms correspond to self-correlations~\cite{DiFrancesco:2016srj,Kitazawa:2017ljq} and to non-dynamical correlations induced by the global conservation law. 
Note that the expression of $d_2$ in Eq.~(\ref{d1234}) is equivalent to Eq.~(\ref{dynamicalvariance}). 

\section{Generalization to net-charge fluctuations}
\label{s:expressions}

The number of positively and negatively charged particles seen in an event are denoted by $N_+$ and $N_-$.
We use the notations $Q$ and $N$ for the net charge and charged multiplicity~\cite{Abelev:2012pv}:
\begin{eqnarray}
  \label{defq}
Q&\equiv&N_+-N_-,\cr
N&\equiv&N_++N_-. 
\end{eqnarray}
One is typically interested in the fluctuations of the net charge, $Q$. 
Ordinary cumulants of net-charge fluctuations are defined by an equation similar to (\ref{defkn}), where $N$ is replaced with $Q$: 
\begin{equation}
 \label{defkn2}
 \ln \left\langle \exp\left(zQ\right)\right\rangle
\equiv\sum_{n=1}^{\infty}\kappa_n\frac{z^n}{n!}.
\end{equation}
These cumulants get trivial contributions to all orders, generated by self-correlations. 
Self-correlations can be removed systematically by constructing factorial cumulants, through a simple generalization of Eq.~(\ref{deffn}): 
\begin{equation}
  \label{deffn2}
 \ln \left\langle  \left(1+z\right)^{N+}\left(1-z\right)^{N-}\right\rangle
\equiv\sum_{n=1}^{\infty}f_n\frac{z^n}{n!}.
\end{equation}
Expressing the left-hand side as a function of $N$ and $Q$ using Eq.~(\ref{defq}) and expanding to order $z^4$, one obtains the explicit expressions~\cite{Bzdak:2016sxg,DiFrancesco:2016srj,Kitazawa:2017ljq}:
\begin{eqnarray}
\label{deff234}
f_1&=&\kappa_1,\cr
  f_2&=& \kappa_2-\langle N\rangle,\cr 
    f_3&=&\kappa_3-3\left(\langle NQ\rangle-\langle N\rangle\langle Q\rangle\right)+2\langle Q\rangle,\cr
  f_4&=&\kappa_4\cr
  &&-6\left(\langle NQ^2\rangle-\langle N\rangle\langle Q^2\rangle
  -2\langle NQ\rangle\langle Q\rangle+2\langle N\rangle\langle Q\rangle^2\right)\cr
  &&+8\left(\langle Q^2\rangle-\langle Q\rangle^2\right)\cr
  &&+3\left(\langle N^2\rangle-\langle N\rangle^2\right)\cr
  &&-6\langle N\rangle.
\end{eqnarray}
Factorial cumulants remove self-correlations order by order, so that $f_n=0$ for $n\ge 2$ for independent particles.
The price to pay is that one also needs the value of $N$ in each event, not just $Q$. This cost is very modest, since $N_+$ and $N_-$ are both measured in every event. 
Self-correlations are not true correlations~\cite{Bzdak:2012ab} and subtracting them does not change the physics one wishes to probe with cumulants.
They are routinely subtracted in analyses of anisotropic flow~\cite{Bilandzic:2010jr,DiFrancesco:2016srj}.
Our point in this paper is that they should also be subtracted in analyses of net-charge fluctuations. 
On the technical side, note that the subtracted terms in Eq.~(\ref{kappa34}) are themselves cumulants~\cite{DiFrancesco:2016srj} of the distribution of $Q$ or $N$ (such as $\langle N^2\rangle-\langle N\rangle^2$) or mixed cumulants involving both $Q$ and $N$.
Note also that $f_3$ ($f_4$)  is odd (even) in $Q$. 
%Factorial cumulants satisfy the same additivity property as ordinary cumulants, in the sense that they add up for independent subsystems.
%If the range of correlations is much smaller than the system size~\cite{Ling:2015yau}, both cumulants and factorial cumulants are proportional to the volume (e.g., the size of the rapidity window where the analysis is carried out).

%point out that in practice if alpha<<1 factorial cumulants are good enough. 
We now define the dynamical cumulants $d_n$ of net charge fluctuations, which generalize factorial cumulants by taking into account the effect of the global charge conservation. This is again done through a simple generalization of Eq.~(\ref{defdn}): 
\begin{equation}
  \label{defdn2}
 \ln \left\langle  \left(1+\frac{e^{\alpha z}-1}{\alpha}\right)^{N+}\left(1+\frac{e^{-\alpha z}-1}{\alpha}\right)^{N-}\right\rangle
\equiv\sum_{n=1}^{\infty}d_n\frac{z^n}{n!}.
\end{equation}
The dynamical cumulants thus defined again interpolate between ordinary cumulants and factorial cumulants: 
If $\alpha=1$, $d_n$ coincides with $\kappa_n$ defined by Eq.~(\ref{defkn2}). In the limit $\alpha\to 0$, $d_n$ coincides with $f_n$ defined by  
Eq.~(\ref{deffn2}). 
Expressing the left-hand side of Eq.~(\ref{defdn2}) as a function of $N$ and $Q$ using Eq.~(\ref{defq}), and expanding to order $z^4$, one obtains the explicit expressions:
\begin{eqnarray}
  \label{defd}
  d_1&=&\kappa_1,\cr
  d_2&=&\kappa_2-(1-\alpha)\langle N\rangle,\cr
  d_3&=&\kappa_3\cr
  &&-3(1-\alpha)\left(\langle NQ\rangle-\langle N\rangle\langle Q\rangle\right)\cr
  &&+(1-\alpha)(2-\alpha)\langle Q\rangle,\cr
  d_4&=&\kappa_4\cr
  &&-6(1-\alpha)\left(\langle NQ^2\rangle-\langle N\rangle\langle Q^2\rangle
  -2\langle NQ\rangle\langle Q\rangle+2\langle N\rangle\langle Q\rangle^2\right)\cr
  &&+4(1-\alpha)(2-\alpha)\left(\langle Q^2\rangle-\langle Q\rangle^2\right)\cr
  &&+3(1-\alpha)^2\left(\langle N^2\rangle-\langle N\rangle^2\right)\cr
  &&-(1-\alpha)(6-6\alpha+\alpha^2)\langle N\rangle.
\end{eqnarray}
The expression of $d_n$ involves the same cumulants of $N$ and $Q$ as the expression of $f_n$, Eq.~(\ref{deff234}), but with different coefficients which depend on $\alpha$. 
Note that $d_n$ can no longer be expressed as a function of ordinary cumulants, as in Eq.~(\ref{d1234}), because it also involves the charged multiplicity $N$ in every event, while $\kappa_n$ only depends on $Q$. 
In the limiting case of one species of particles ($N_-=0$, which implies $N=Q$), Eqs.~(\ref{defd}) reduce to Eqs.~(\ref{d1234}), as they should. 

By construction, dynamical cumulants of order $\ge 2$ eliminate contributions of binomial fluctuations:
If $N_+$ and $N_-$ follow independent binomial distributions with the same fraction $\alpha$, using Eq.~(\ref{dynbinomial}), one obtains 
\begin{equation}
\label{dynbinomial2}
\left\langle  \left(1+\frac{e^{\alpha z}-1}{\alpha}\right)^{N+}\left(1+\frac{e^{-\alpha z}-1}{\alpha}\right)^{N-}\right\rangle=
\exp(\langle Q\rangle z),
\end{equation}
where $\langle Q\rangle=\alpha(N_+^{\rm tot}-N_-^{\rm tot})$ is the average net charge $Q$ seen in the detector. 
Inserting into Eq.~(\ref{defdn2}), this implies $d_n=0$ for $n\ge 2$. 
 
By contrast, ordinary cumulants do not vanish. 
If $N_+$ and $N_-$ follow binomial distributions, a simple calculation gives
\begin{equation}
\ln\left\langle e^{\pm z N_\pm}\right\rangle=\frac{\langle N_\pm\rangle}{\alpha}\ln\left(\alpha e^{\pm z}+1-\alpha\right). 
\end{equation}
Expanding in powers of $z$ and using Eq.~(\ref{defkn2}), one obtains the expressions~\cite{Braun-Munzinger:2018yru}:  
\begin{eqnarray}
\label{consk123}
\kappa_1&=&\langle Q\rangle,\cr
\kappa_2&=&(1-\alpha)\langle N\rangle,\cr
\kappa_3&=&(1-\alpha)(1-2\alpha)\langle Q\rangle,\cr
\kappa_4&=&(1-\alpha)(1-6\alpha+6\alpha^2)\langle N\rangle.
\end{eqnarray}
Odd-order cumulants are proportional to the mean charge, while even-order cumulants are proportional to the charged multiplicity. 
In the limit $\alpha\to 0$, $\kappa_3=\kappa_1$ and $\kappa_4=\kappa_2$. 
A finite $\alpha$ results in reduced higher-order cumulants, $\kappa_3<\kappa_1$ and $\kappa_4<\kappa_2$. 

\section{Volume fluctuations}
\label{s:volume}

Dynamical cumulants are by construction insensitive to the correlation induced by the global conservation law. 
We now discuss their sensitivity to impact parameter fluctuations in a centrality bin, which is another non-dynamical fluctuation. 
More specifically, we assume that the distribution of $N_+$ and $N_-$ are binomial distribution at a fixed impact parameter, but the centrality bin consists of a range of impact parameters. 

One can carry out the average over events in Eq.~(\ref{defdn2}) in two steps:
First, one averages over events with the same impact parameter. 
Since the distributions of $N_+$ and $N_-$ are assumed binomial, Eq.~(\ref{dynbinomial2}) holds, where $\langle Q\rangle $ is a function of $b$ which we denote by ${\cal Q}(b)$. 
Second, one averages over impact parameter:
\begin{equation}
\label{fqb}
\sum_{n=1}^{\infty}d_n\frac{z^n}{n!}=  \ln\langle\exp\left(z {\cal Q}(b)\right)\rangle,
\end{equation}
where angular brackets in the right-hand side now denote the average over impact parameter, while ${\cal Q}(b)$ is already averaged over events at fixed $b$. 

At LHC energies and top RHIC energies, where the system is nearly charge symmetric~\cite{Abelev:2012wca}, ${\cal Q}(b)\approx 0$ for all $b$, therefore dynamical cumulants vanish to all orders in the absence of dynamical fluctuations, even if the analysis is done in a wide centrality bin. 

At lower RHIC energies~\cite{Adamczyk:2014fia,Adare:2015aqk}, there is a significant charge asymmetry and ${\cal Q}(b)$ no longer vanishes. 
If ${\cal Q}(b)$ varies with $b$ and if there is a range of values of $b$ in the centrality bin, $d_n$ no longer vanishes for $n\ge 2$. 
This non-zero value solely results from the variation of the {\it mean\/} charge ${\cal Q}(b)$ with $b$. 
More specifically, $d_n$ is the cumulant of order $n$ of the distribution of ${\cal Q}(b)$ in the centrality bin:
\begin{eqnarray}
\label{cumulnb}
d_1&=& \langle {\cal Q}(b)\rangle,\cr
d_2&=& \left\langle ({\cal Q}(b) -\langle {\cal Q}(b)\rangle)^2\right\rangle,\cr
d_3&=& \left\langle ({\cal Q}(b) -\langle {\cal Q}(b)\rangle)^3\right\rangle.
\end{eqnarray}
For instance, $d_2$ is the variance of ${\cal Q}(b)$ due to impact parameter fluctuations. 
Hence, impact parameter fluctuations generate $d_2>0$.
However, this effect is numerically small, as will be shown explicitly in Sec.~\ref{s:mc}.
If $\delta {\cal Q}$ denotes the width of the distribution of ${\cal Q}(b)$ in the centrality bin, which is typically proportional to the width of the centrality bin itself, then the contribution of centrality fluctuations to $d_n$ is proportional to $(\delta {\cal Q})^n$: In practice, dynamical cumulants of order 3 and higher are essentially insensitive to centrality fluctuations. 
By contrast, ordinary cumulants are strongly sensitive to volume fluctuations, as will be shown in Sec.~\ref{s:mc}.

The same reasoning can be applied to factorial cumulants by taking the limit where the acceptance fraction $\alpha$ goes to 0, i.e., replacing the binomial distribution (\ref{binomial}) with a Poisson distribution, and $d_n$ [defined by Eq.~(\ref{defdn2})] with $f_n$ [defined by Eq.~(\ref{deffn2})]. 
One concludes that if $N_+$ and $N_-$ follow Poisson distributions at fixed impact parameter, and if the system is charge symmetric, then factorial cumulants vanish to all orders, irrespective of the width of the centrality bin. If there is a charge asymmetry, the sensitivity to volume fluctuations is still greatly reduced compared to ordinary cumulants, and the reduction is more important for higher-order cumulants. 

\section{Resonance decays}
\label{s:resonance}

We now briefly discuss the effect of resonance decays~\cite{Nahrgang:2014fza}. 
Decays do not modify conserved charges by definition.
Hence, if all decay products are recorded in the detector, the cumulants, which solely depend on the conserved charge [Eq.~(\ref{defkn2})], are strictly unchanged after the decay. 
Factorial cumulants, however, are in general modified, and so are the dynamical cumulants.
We evaluate this modification on the example of the decay $\rho^0\to\pi^+\pi^-$. 
We assume for simplicity that all charged particles originate from $\rho^0$ decays, and that the decay products are all detected. 
We first consider the case where the number $N_0$ of $\rho^0$ mesons follows a Poisson distribution, corresponding to uncorrelated emission:
\begin{equation}
  \label{poisson}
P_{N_0}=\frac{\langle {N_0}\rangle^{N_0}}{{N_0}!}e^{-\langle {N_0}\rangle}.  
\end{equation}
Each $\rho^0$ gives one $\pi^+$ and one $\pi^-$, hence $N^+=N^-={N_0}$, and the total charged multiplicity is $N=2N_0$. 
The generating function of factorial cumulants, given by Eq.~(\ref{deffn2}), can be easily evaluated:
\begin{eqnarray}
  \label{reso}
  \left\langle  \left(1+z\right)^{N+}\left(1-z\right)^{N-}\right\rangle
  &=&\sum_{{N_0}=0}^{\infty}(1+z)^{N_0}(1-z)^{N_0} \frac{\langle {N_0}\rangle^{N_0}}{{N_0}!}e^{-\langle {N_0}\rangle}\cr
  &=&\exp\left(-\langle {N_0}\rangle z^2\right). 
\end{eqnarray}
Inserting into Eq.~(\ref{deffn2}), one obtains
\begin{eqnarray}
\label{reso2}
  f_1&=&0\cr
f_2&=&-2\langle {N_0}\rangle=-\langle N\rangle\cr
f_3&=&f_4=\cdots=0. 
\end{eqnarray}  
Before the decays, there is no charged particle, and all factorial cumulants are 0. 
Eq.~(\ref{reso2}) shows that the only factorial cumulant which is modified by the resonance decay is the cumulant of order 2.
The physical explanation of this result is that a two-body decay contributes to the 2-particle correlation, which is measured by $f_2$, but not to higher-order correlations. 
Kitazawa {\it et al.\/}~\cite{Kitazawa:2017ljq} write that ``factorial cumulants would be altered in non-trivial ways'' by decay processes, but our conclusion also applies to the decays of a doubly-charged particle into two singly- charged particles considered in their paper.
The calculation above can be easily generalized: A three-body decay modifies factorial cumulants up to order three, etc. 

This exact cancellation of higher-order cumulants holds for factorial cumulants, not for dynamical cumulants. 
In the example above, $d_3$ still vanishes by symmetry, not $d_4$. 
Repeating the algebra with Eq.~(\ref{defdn2}) instead of Eq.~(\ref{deffn2}), or using the explicit expressions Eq.~(\ref{defd}) with $Q=0$ and $N=2N_0$, one finds $d_4=\alpha^2 d_2$.
Since $\alpha$ is typically much smaller than unity, as will be pointed out in Sec.~\ref{s:mc}, one sees that the spurious $d_4$ produced by resonance decays is small.
%Generally, one expects that a $n$-body decay contributes little to dynamical cumulants of order larger than $n$. 

\section{Monte Carlo simulations}
\label{s:mc}

%for phenix we make the approxiamtion alpha<<1 so that dynamical cumulants reduce to factorial cumulants. 

We now illustrate effects of non-dynamical fluctuations  by means of realistic simulations. 
We show that non-dynamical fluctuations alone are likely to explain results obtained at RHIC by the STAR~\cite{Adamczyk:2014fia} and the PHENIX~\cite{Adare:2015aqk} Collaborations for the fluctuations of the net electric charge.
We show that if dynamical cumulants were used instead of ordinary cumulants, effects of non-dynamical fluctuations would be largely suppressed. 

\begin{figure*}[t!]
\begin{center}
\includegraphics[width=.84\linewidth]{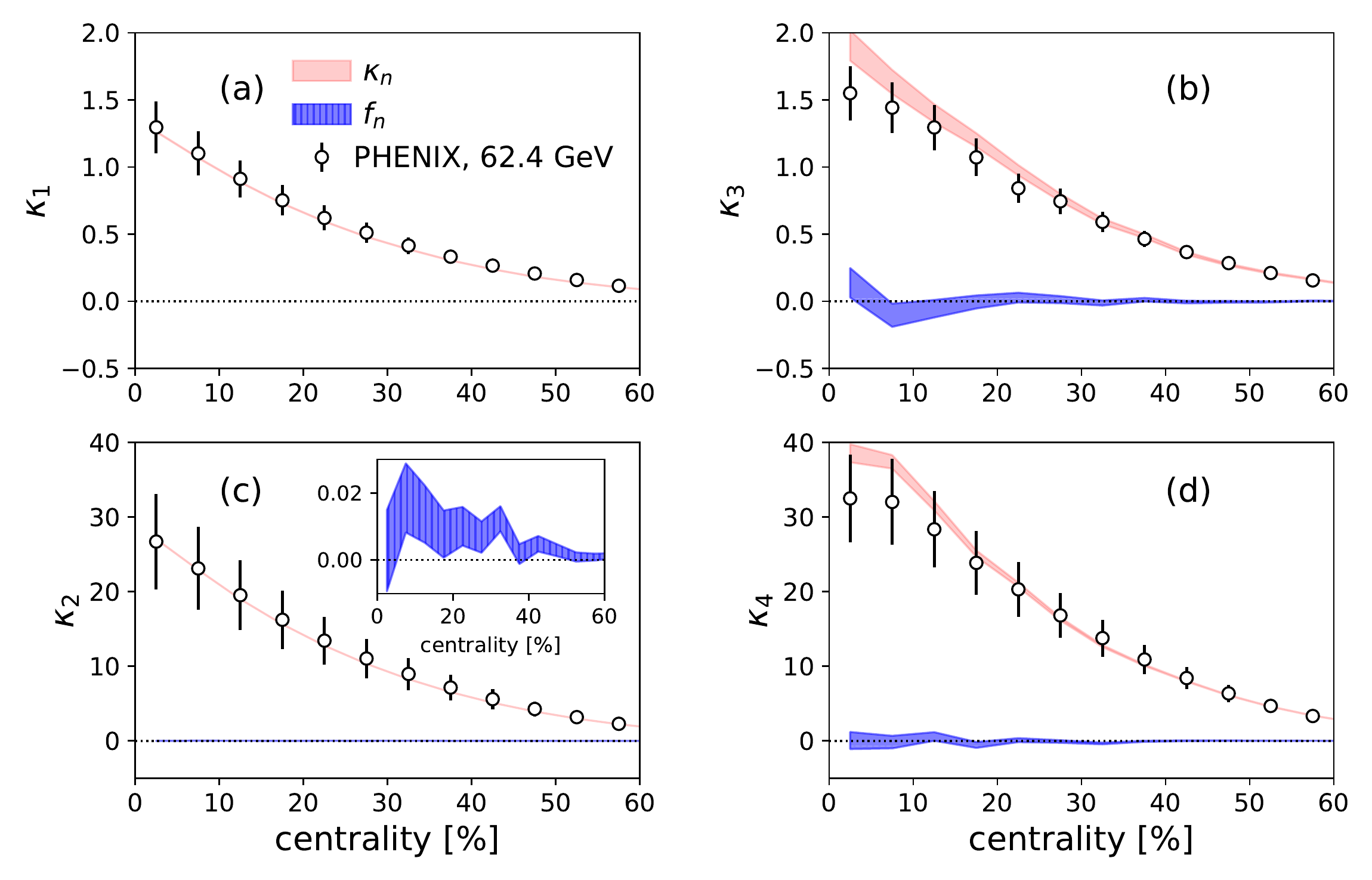} 
\end{center}
\caption{(Color online) 
\label{fig:phenix62}
Cumulants of the net charge distribution in Au+Au collisions at $\sqrt{s_{\rm NN}}=62.4$~GeV as a function of centrality percentile. 
Symbols: PHENIX data~\cite{Adare:2015aqk} with errors (systematic only). Bands: Results of our Monte Carlo simulations for the usual net-charge cumulants ($\kappa_n$) and for the factorial cumulants ($f_n$) defined by Eq.~(\ref{deff234}). The width of the bands corresponds to the statistical error of our Monte Carlo simulation. The dynamical cumulants $d_n$ coincides with $f_n$ since $\alpha\ll 1$. 
(a) $n=1$ (mean); 
(b) $n=3$ (skewness); 
(c) $n=2$ (variance). The inset is a zoom showing the value of $f_2$; 
(d) $n=4$ (kurtosis). 
}
\end{figure*}
We assume that the distributions of $N_+$ and $N_-$ are binomial distributions at fixed impact parameter $b$, so that there are no dynamical fluctuations. 
The only other source of fluctuation in our simulation is the fluctuation of impact parameter within a centrality bin. 
Experimental analyses of net charge fluctuations are carried out typically in 5\% centrality bins. 
The experimental definition of the centrality uses an observable $n$ (charged multiplicity or energy seen in a specific detector), whose relation with $b$ is not one to one. 
The probability distribution of $b$ in a centrality bin can be reconstructed using the measured histogram of $n$ under fairly general assumptions~\cite{Das:2017ned,Rogly:2018ddx}. 
The distribution of $n$ is rarely made public, so that we do not have this information for all energies and experiments. 
Throughout this article, we use STAR data for Au+Au collisions at $\sqrt{s_{\rm NN}}=130$~GeV~\cite{Adler:2001yq}. 
We assume that the distribution of $b$ would not change drastically from STAR data to PHENIX data, or as one varies the collision energy $\sqrt{s_{\rm NN}}$. 

The Monte Carlo simulation is done  in the following way. 
For each event, we draw randomly the true centrality (defined according to impact parameter, and denoted by $c_b$~\cite{Das:2017ned}) between 0 and 1. 
We then draw independently the values of $N_+$, $N_-$ and $n$. 
We assume that $N_+$ and $N_-$ follow binomial distributions, while $n$ follows a gamma distribution~\cite{Rogly:2018ddx}.
The parameters of the gamma distribution are determined by the mean and the variance of the distribution of $n$ at fixed impact parameter, which are given by Table II of Ref.~\cite{Das:2017ned}. 
We assume, for sake of simplicity, that the mean values of $N_+$ and $N_-$ at a given impact parameter are equal to the mean value of $n$, up to a global proportionality factor. 
Thus, the only free parameters in our calculation are the mean values of $N_+$ and $N_-$ for central collisions ($b=0$), which depend on the detector and on the collision energy, and the fraction of particles falling in the detector acceptance, $\alpha$. 
We generate a large number of events, and then classify them into centrality bins according to the value of $n$, as in experiments. 
We then evaluate the cumulants $\kappa_n$ and the dynamical cumulants $d_n$, with $n=1,\cdots,4$ as explained in Sec.~\ref{s:expressions}.

Figure~\ref{fig:phenix62} displays PHENIX data~\cite{Adare:2015aqk} for $\sqrt{s_{\rm NN}}=62.4$~GeV together with the results of our Monte Carlo simulations. We generate $2\times 10^8$ events. 
Since the acceptance of the PHENIX analysis covers only half of the range in azimuth, and a small interval in pseudorapidity, the fraction of particles detected $\alpha\ll 1$. 
We therefore carry out this simulation in the limit $\alpha\to 0$. In this limit, $N_+$ and $N_-$ follow Poisson distribution. 
The other free parameters in our calculation are the mean values of $N_+$ and $N_-$ for central collisions. We fix them to the values $15.7$ and $14.3$, respectively so as to match the measured values of $\kappa_1$ and $\kappa_2$ for central collisions.  
As shown in Fig.~\ref{fig:phenix62}, we then explain the values of $\kappa_1$ and $\kappa_2$  for other centralities, as well as the values of $\kappa_3$ and $\kappa_4$, without any additional parameter. 
In particular, we explain the observation that $\kappa_3>\kappa_1$ and  $\kappa_4>\kappa_2$, at variance with the result for the binomial distribution (\ref{consk123}). 
This effect is produced by the fluctuations of impact parameter in a centrality bin. 
Note that $\kappa_3$ and $\kappa_4$ are slightly overpredicted in central collisions, but given the crudeness of our model, we consider that agreement is satisfactory. 
Our results imply that even though experimental results are seemingly non trivial, they can be explained without invoking dynamical fluctuations. 
We also evaluate the dynamical cumulants, which reduce here to factorial cumulants, $d_n=f_n$, since we take the limit $\alpha\to 0$. 
They are essentially compatible with zero, which illustrates our statement, in Sec.~\ref{s:volume}, that volume fluctuations do not artificially generate dynamical cumulants. 
Zooming in (inlay in Fig.~\ref{fig:phenix62} (c)), one sees that $f_2$ is actually positive, as expected from the discussion in Sec.~\ref{s:volume}.
 But it is smaller than the ordinary cumulant $\kappa_2$ by a factor 1000.  
Note that short-range correlations due to resonance decays can have a sizable effect on $f_2$, but not on $f_3$ and $f_4$, as discussed in Sec.~\ref{s:resonance}. 

\begin{figure*}[t!]
\begin{center}
  \includegraphics[width=.84\linewidth]{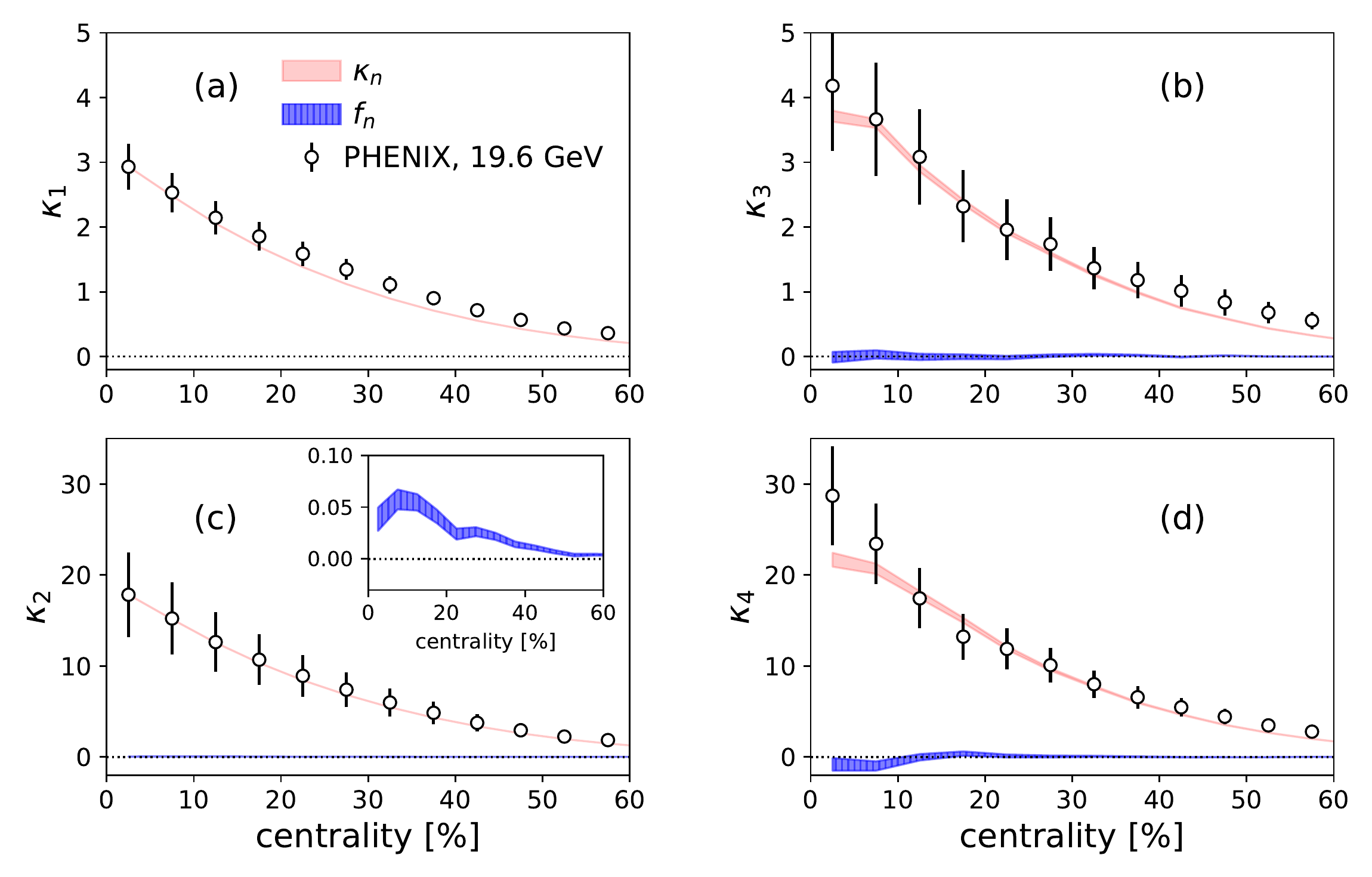} 
\end{center}
\caption{(Color online) 
\label{fig:phenix19}
Same as Fig.~\ref{fig:phenix62} for Au+Au collisions at $\sqrt{s_{\rm NN}}=19.6$~GeV. 
Data from the PHENIX Collaboration~\cite{Adare:2015aqk}. 
}
\end{figure*}
In Fig.~\ref{fig:phenix19}, we repeat the exercise at the lower energy $\sqrt{s_{\rm NN}}=19.6$~GeV. The charge asymmetry is larger at lower energy, hence $\kappa_1$ is larger than in Fig.~\ref{fig:phenix62}. 
The charged multiplicity, on the other hand, is smaller at the lower energy, so that $\kappa_2$ is smaller. 
In our simulation, we again assume that $\alpha\approx 0$, so that dynamical cumulants reduce to factorial cumulants. 
The mean values of $N_+$ and $N_-$ in central collisions are fixed to the values $11.5$ and $8.25$, respectively. 
We generate $10^8$ events. 
As in Fig.~\ref{fig:phenix62}, we reproduce the data without dynamical fluctuations. 
Factorial cumulants are again close to 0. 
The factorial moment $f_2$ is slightly larger, which is the modest consequence of the larger charge asymmetry: 
More quantitatively, the charge seen in the detector increases by a factor 2 (compare panels (a) of Fig.~\ref{fig:phenix62} and Fig.~\ref{fig:phenix19}), and $f_2$ is proportional to the variance of the charge according to the discussion in Sec.~\ref{s:volume}, therefore, it increases by a factor $\sim 4$.

\begin{figure*}[t!]
\begin{center}
\includegraphics[width=.84\linewidth]{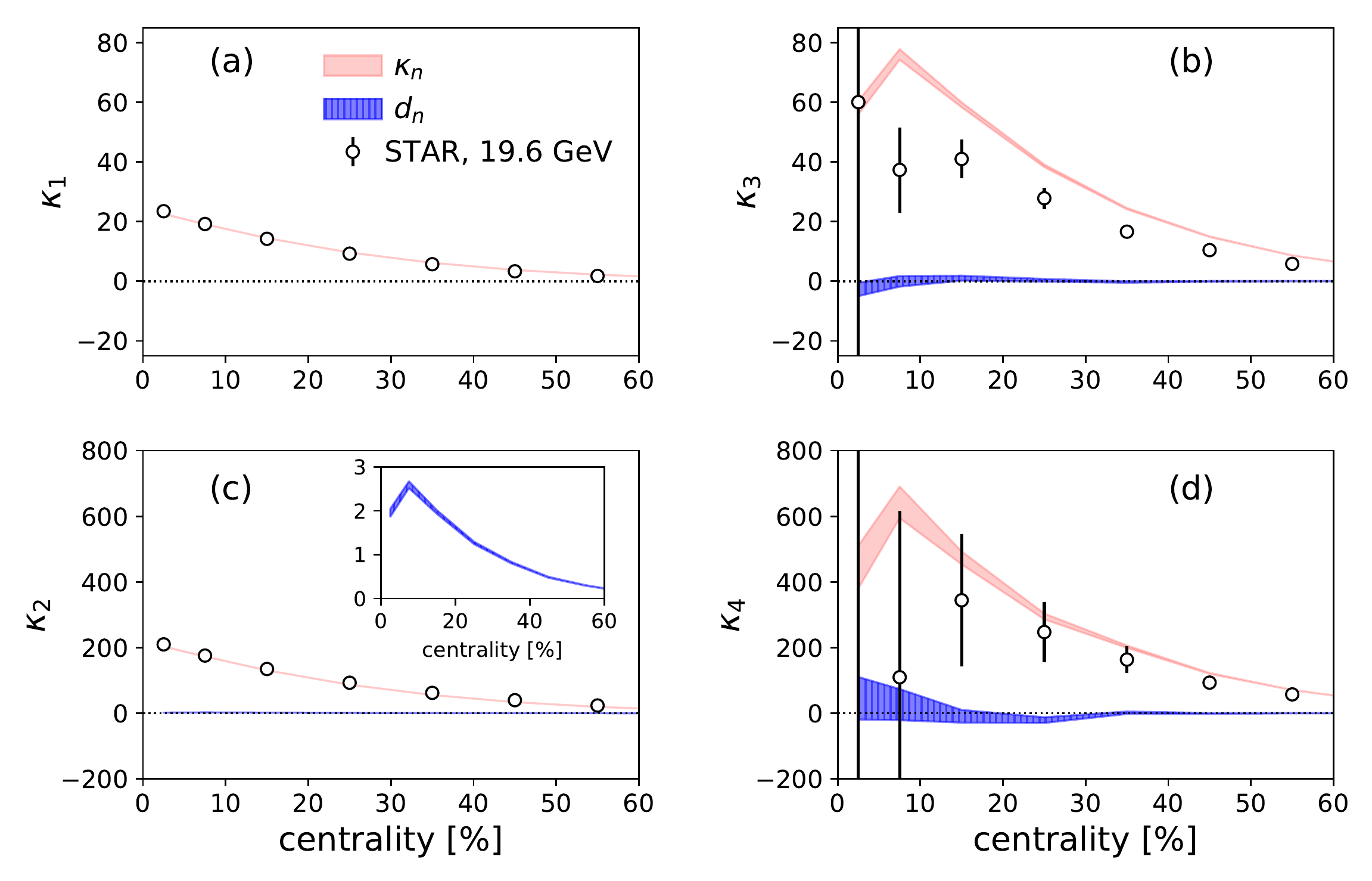} 
\end{center}
\caption{(Color online) 
\label{fig:star19}
Same as Fig.~\ref{fig:phenix19} using STAR data~\cite{Adamczyk:2014fia} for Au+Au collisions at $\sqrt{s_{\rm NN}}=19.6$~GeV. 
Instead of the factorial cumulants $f_n$, we now evaluate the dynamical cumulants defined by Eqs.~(\ref{defd}), with $\alpha=0.216$ (see text). 
As in Fig.~\ref{fig:phenix19}, the inset in panel (c) is a zoom showing the value of $d_2$. 
}
\end{figure*}
We now discuss STAR data at the same energy $\sqrt{s_{\rm NN}}=19.6$~GeV~\cite{Adamczyk:2014fia}. 
The main difference with PHENIX data is the larger acceptance in pseudorapidity and azimuth: the detector sees more particles. 
Using the pseudorapidity distribution published by STAR at the same energy~\cite{Adams:2005cy}, we estimate that the acceptance fraction is $\alpha\approx 0.216$. 
The mean values of $N_+$ and $N_-$ in central collisions are chosen to be $155$ and $130$, respectively,\footnote{Note that the relative charge asymmetry is smaller than that seen by PHENIX at the same energy. We have not investigated the origin of this difference.} 
in order to reproduce the magnitudes of $\kappa_1$ and $\kappa_2$. 
We generate $2\times 10^8$ events. 
%The larger multiplicities explain the larger vertical scales, compared to PHENIX data ($\kappa_n$ is typically proportional to the acceptance). 
A technical difference with the PHENIX analysis is that the STAR analysis is first carried out in 1\% centrality bins, which are then recombined into 5\% or 10\% bins. 
We follow the same procedure in our simulation, whose results are shown in Fig.~\ref{fig:star19}. 
The use of narrower bins reduces the effect of impact parameter fluctuations. 
However, they still have a significant effect~\cite{Braun-Munzinger:2018yru}: 
As in the case of PHENIX data, the orderings $\kappa_3>\kappa_1$ and $\kappa_4>\kappa_2$, which could naively be attributed to dynamical fluctuations, are reproduced (and even overpredicted) by our calculation.
Our model reproduces the cumulant ratio $\kappa_4/\kappa_2$ better than the model used by the STAR Collaboration~\cite{Adamczyk:2014fia}, where $N_+$ and $N_-$ are assumed to be independent and to follow negative binomial distributions.
The reason why we achieve a better description is that we take into account the correlation between $N_+$ and $N_-$ induced by impact parameter fluctuations within a centrality bin. 
As in the case of PHENIX data, the dynamical cumulant $d_2$ is slightly positive (inlay in Fig.~\ref{fig:star19} (c)), but smaller than the ordinary cumulant $\kappa_2$ by a factor $\sim 100$. 
Higher-order dynamical cumulants $d_3$ and $d_4$ are compatible with zero.
Any effect of resonance decays on these results would be within the present error bars, according to the estimates in Sec.~\ref{s:resonance}. 
These results illustrate that dynamical cumulants are insensitive to non-dynamical fluctuations.

\section{Conclusions}

We have shown that existing data on net-charge fluctuations show no clear evidence of dynamical fluctuations.
As long as one uses cumulants to characterize net-charge fluctuations, the observation of dynamical fluctuations will be hindered by the large effects of volume fluctuations. 
One knows how to subtract their effects for two-particle correlations~\cite{Olszewski:2017vyg}, but not for higher-order correlations.
If volume fluctuations are not under control, comparison with lattice data, where the volume is fixed, is ambiguous.

%we have introduced new observables... 
We have introduced new observables, called dynamical cumulants, whose expressions are given in Eq.~(\ref{defd}). 
They generalize factorial cumulants by taking into account the correlation due to global charge conservation. 
The dynamical cumulant of order $n$ contains the same information about the $n$-particle correlation as the ordinary cumulant $\kappa_n$. 
These two quantities differ only by terms induced by self-correlations and global charge conservation. 
Dynamical cumulants  offer the same flexibility and can also be analyzed in various rapidity windows~\cite{Brewer:2018abr}.
We have shown that unlike cumulants, factorial cumulants and dynamical cumulants are remarkably insensitive to volume fluctuations. 
They cannot be directly compared with lattice data, but can be used to obtain direct and reliable information on interactions and dynamical correlations among the produced hadrons, which in turn can be used in comparing with ab-initio calculations~\cite{Huovinen:2017ogf}. 

\section*{Acknowledgments}
This research was supported by the Munich Institute for Astro- and Particle Physics (MIAPP) of the DFG cluster of excellence ``Origin and Structure of the Universe''.

\end{document}